\documentclass[letterpaper]{sig-alternate}

\usepackage{graphicx, subfigure}
\usepackage{algorithm}
\usepackage{algpseudocode}
\usepackage{amsmath}
\usepackage{tabularx}
\usepackage{url}
\usepackage{paralist}
\usepackage[colorlinks, citecolor=black, urlcolor=black, linkcolor=black]{hyperref}
\usepackage{flushend}

\usepackage{ifpdf}
\ifpdf
\else
\fi

\DeclareMathOperator*{\argmax}{arg \, max}

\DeclareMathOperator*{\cov}{cov}
\newcommand*\tageq{\refstepcounter{equation}\tag{\theequation}}
\newcommand{\bs}{\boldsymbol}

\begin{document}
\conferenceinfo{CIKM'12,} {October 29--November 2, 2012, Maui, HI, USA.} 
\CopyrightYear{2012} 
\crdata{978-1-4503-1156-4/12/10} 
\clubpenalty=10000 
\widowpenalty = 10000


\title{Sequential Selection of Correlated Ads by POMDPs}
\numberofauthors{1} \author{
\alignauthor
	Shuai Yuan, Jun Wang\\
       \affaddr{Department of Computer Science, University College London}\\
       \email{\{s.yuan, j.wang\}@cs.ucl.ac.uk}
}

\maketitle

\begin{abstract} 

Online advertising has become a key source of revenue for both web search engines and online publishers. For them, the ability of allocating \emph{right} ads to \emph{right} webpages is critical because any mismatched ads would not only harm web users' satisfactions but also lower the ad income. In this paper, we study how online publishers could optimally select ads to maximize their ad incomes over time. The conventional offline, content-based matching between webpages and ads is a fine start but cannot solve the problem completely because good matching does not necessarily lead to good payoff.  Moreover, with the limited display impressions, we need to balance the need of selecting ads to learn true ad payoffs (exploration) with that of allocating ads to generate high immediate payoffs based on the current belief (exploitation). In this paper, we address the problem by employing Partially observable Markov decision processes (POMDPs) and discuss how to utilize the correlation of ads to improve the efficiency of the exploration and increase ad incomes in a long run. Our mathematical derivation shows that the belief states of correlated ads can be naturally updated using a formula similar to collaborative filtering. To test our model, a real world ad dataset from a major search engine is collected and categorized.  Experimenting over the data, we provide an analyse of the effect of the underlying parameters, and demonstrate that our algorithms significantly outperform other strong baselines.  \end{abstract}




\section{Introduction}
Online advertising has received blooming development in recent years. The volume of the industry has grown from \$8.1 billion in 2000 to \$26 billion in 2010 with a new record every year \cite{iab2011}. Online advertising is vitally important for both web search engines and online content providers and publishers because it provides them with major sources of revenue. For example, Google search engine was serving 7.2 billion page views per day in 2011 \cite{jeff2011amazing} and its revenue is mainly driven by advertising products like AdWords, DoubleClick, and AdMob. Similarly, online content providers such as Yahoo! and news agencies such as New York Times increasingly choose advertising as the smarter alternative to conventional subscription based services.

Online publishers can make profit by selling \textit{impressions} (an instance of an ad being seen on users' monitors). A fundamental problem faced by publishers is that given limited display impressions and ad slots, how to maximize their ad incomes over time. It is a complex problem because in practice publishers usually need to make decisions considering various aspects. First, they have to decide whether to deal private contracts with advertisers or agencies directly, or to participate in public ad networks or exchanges to reach more demands. In \cite{fridgeirsdottir2007revenue} a queuing system was proposed to find the optimal policy of selecting advertisers to make private contracts, whereas in \cite{bharadwaj2010pricing, li2009pricing}, the focus is on pricing ads properly in either of the two settings. Second, the payment scheme is different and the publishers need to choose between pay-per-view, pay-per-click, and possibly other models. A balance can be established in a static setting as reported in \cite{mangani2004online, fjell2009online, kwon2011single}. Lastly and most importantly, a decision has to be made on which page \cite{broder2007semantic, wu2011leveraging} and which users \cite{tyler2011retrieval, tang2011learning} these ads should be matched with, probably in a real-time fashion using text summaries \cite{anagnostopoulos2007just}. Traditionally the matching is done by ad networks and advertisers are allowed to choose which keywords they intend to bid. Now publishers are getting more involved and can actively switch in real-time between different pricing schemes and different networks in order to increase their ad incomes, as demonstrated in the Google DFP Small Business system \cite{google-double-clicks} and some ad exchanges like AdBrite (\url{http://www.adbrite.com/}).

This essentially allows online publishers to integrate the above decisions all together in a more general framework to optimize their ad revenue. Yet, the following challenges remain unsolved because of its dynamic settings: First, the prior analysis of the matching between content and ads provides a fine start but a good payoff is not necessarily guaranteed by a good matching and need to be verified over time.  With the limited display impressions, we also need to balance the need of selecting ads and their pricing schemes to learn the true payoffs (exploration) with that of allocating ads to generate high immediate payoffs based on the current belief (exploitation). Second, online publishers have more opportunities to explore from different ad networks and pricing schemes provided ads are correlated.  A key question is how to make use of the correlations embedded in the data to improve the efficiency of the exploration and increase the ad incomes in a long run.  In this paper, we study the above two issues and formulate the sequential ad selection problem by applying Partially observable Markov decision processes (POMDPs).  To provide a basic understanding of publisher revenue problem when dealing with multiple ads or ad sources, we do not distinguish between various trade mechanisms and pricing models, but instead consider them essentially the same and focus on their payoffs. Moreover, to make our study focused, we formulate our problem by considering the correlation of ads, while bearing in mind that the same principle and result can be applied to the correlation of webpages and users. Our mathematical derivation shows that the belief states of correlated ads can be naturally updated using a formula similar to collaborative filtering. We examine the model on a carefully collected dataset, and the results show that our models, particularly with the new belief updates, outperform other strong baselines.

The remainder of the paper is organized as follows: in Section~\ref{sec-related-work} we provide a discussion about related work and in Section~\ref{sec-problem-statement} we present the proposed model and its approximate solutions. The experiments are given in Section~\ref{sec-experiment} and conclusions are presented in Section~\ref{sec-conclusion}.

\section{Related work}
\label{sec-related-work}

Previous research on publisher revenue optimization has been mainly focused on display ads and their (private) contracts. Traditionally the research question is limited to how to choose advertisers to make (private) contracts with, as elaborated in \cite{chickering2003targeted, fridgeirsdottir2007revenue, roels2009dynamic}. In \cite{fridgeirsdottir2007revenue} the authors used a queueing system to accommodate advertisers and added constraint that advertisers are impatient and would leave if their ads cannot be displayed right way. A dynamic programming solution was provided in \cite{roels2009dynamic} by combining the available ad inventories and dynamically delivery of promised advertising contract to the viewers.  Although our model also optimizes publisher revenue over time, the settings are different: we focus on contextual advertising, where ads can be specified using keywords or categories, and study how the matching between webpages and ads can be established w.r.t. payoffs over time. We introduce a simple yet general dynamical payoff model that helps to address the inventory management problem while taking into account the correlation of ads. 
 
An interesting variation of ad inventory management problem is to find the optimal pricing models. The pay-per-view and pay-per-click models were discussed in \cite{hu2010pricing}.  In \cite{mangani2004online}, the study concluded that a combination of pay-per-view and per-per-click pricing models may be the most optimal for an advertiser who has to choose between them. The same choice problem has later been studied in \cite{fjell2009online} and the author concluded that the optimal choice for a price taking publisher is either pay-per-view or per-per-click but not a combination of those two pricing models. Recently similar conclusion was reported in \cite{kwon2011single}, in which the competition between two models was discussed but limited to a static setting: a single period of planning horizon. Similarly in \cite{fridgeirsdottir2007revenue} the authors assumed that the publisher could choose ad networks to utilize remnant impressions after fulfilling contracts. Our goal is similar to those papers since we are also targeted to find optimal ad sources. However, our model is focused on real-time impression-based setting; we also utilize the correlation of ads to accelerate the discovery process, and provide an analyse of the optimization problem over time. For instance, different from \cite{fjell2009online} we do not make any assumption on market elasticity; neither do we constraint the publisher to a contract-first setup \cite{fridgeirsdottir2007revenue}.

Another aspect of publisher revenue optimization is ad scheduling with the constraint of geometrical features of website, uncertainty of advertisers and available ad inventory (impressions) \cite{kumar2006scheduling, roels2009dynamic, turner2011scheduling}. If the publisher fails to deliver promised impressions a good-will penalty would be incurred. In \cite{turner2011scheduling} the ad scheduling problem was modelled in video games and a dynamic algorithm was developed.  When considering pricing models, e.g. pay-per-click, publishers need to understand the webpage content and user preferences to achieve a better scheduling. Because now displaying ads may not earn the publisher anything if ads are not relevant or users are not interested therefore no clicks. Traditionally optimizing matching between webpages and ads belongs to the field of contextual advertising research \cite{broder2007semantic, ribeiro2005impedance}. In \cite{broder2007semantic} a system utilizing both semantic and syntactic features was proposed to address the problem. Similarly the correlation of user-ad (behaviour targeting) has also been studied previously in \cite{wang2002understanding, chakrabarti2008contextual}, focusing on improving advertising effectiveness by modelling attitude and feedback from users. Our study can be consider as a dynamical extension of the above works. The contextual matching can be used as a prior belief in our model and it can be updated and verified using our update formulas sequentially. 

In terms of techniques, our work is closely related to multi-arm bandit with dependent arms discussed in \cite{pandey2007multi} as we also face a multi-period selection problem with an exploration and exploitation dilemma. In \cite{pandey2007multi} the dependencies of candidates are modelled by clustering them first; after that, a multi-arm bandit algorithm runs twice: first selects a cluster and then a candidate in that cluster. Our paper is different in that we directly model the dependencies using a covariance matrix. The impact of correlation is well formulated and illustrated by rigorously deriving the belief updates once other correlated ads have been selected. Specifically, the problem is formulated by applying Partially Observable Markov Decision Processes (POMDPs) with discrete action (selection of ads), continuous observations (payoffs), and continuous hidden states (performance of ads). Our model is a special case of continuous POMDPs \cite{cassandra1998survey, thrun2000monte}, where two-stage Gaussian generative processes and no transit of hidden states are considered during the planning horizon. To provide an optimal ad selection, similar with \cite{thrun2000monte, spaan2005perseus}, we follow the Monte Carlo sampling approach to deal with continuous observations and to approximate a Dynamic Programming solution in the finite planning horizon. As a complement of the approximation approach, we also extend an Upper Confidence Bound algorithm \cite{auer2002finite} by integrating it with our belief update method.

\newpage

\section{The proposed model}
\label{sec-problem-statement}

\subsection{The Sequential Payoff Model}
In this section we formulate the sequential ad selection problem for a publisher-side system. Suppose there is an online publisher who wants to put ads on their hosting webpages to generate profits. The ads could be obtained from various sources either by making contracts with advertisers directly, by registering with ad networks, or by employing a supply-side-platform \cite{google-double-clicks}.  We now formally introduce the sequential payoff model to describe how the publisher revenue is earned from ads. Suppose there are $N$ ads from various ad networks or exchanges available for the publisher. For each display impression, the publisher needs to decide which ad should be selected. Without loss of generality, we consider to make our selection decisions every $M$ impressions and denote the decision times as $t\in [1,...,T]$.  To stay focused, we consider the impression has only one ad slot, while bearing in mind that the scenario of multiple ad slots per impression can be addressed by incorporating user click-through models (for pay-per-click ads) to remove rank bias \cite{Richardson:2007} or consider them from different impressions (for pay-per-view ads).

For each time step $t\in [1,...,T ]$, we define 
\begin{equation}
\Psi(t) = \begin{bmatrix}
	s(1), & x(1) \\
	\dots, & \dots \\
	s(t-1), & x(t-1)
\end{bmatrix}
\end{equation}
as the available information up to time $t$, where $s(t) \in N$ denotes the decision, i.e. the index of ad selected for the time step $t$. We use the random variable $X(t)$ to denote the payoff gained at time step $t$ and $x(t)$ its realization.

Let $\pi$ be an arbitrary choosing policy according to the information obtained so far. We define
\begin{equation}
s(t)  \equiv \pi ( \Psi(t) )  
\end{equation}
To simplify our notation, we use them exchangeably in the rest of the paper. The cumulative payoff over time $T$ with certain policy $\pi$ is
\begin{equation}\label{objectfunction}
R_{\pi}(T) \equiv M \sum_{t=1}^T  X_{s(t)}(t)
\end{equation}

\begin{figure}[t]
	\centering
	\includegraphics[width=0.45 \textwidth, bb=0 0 4.68in 2.42in]{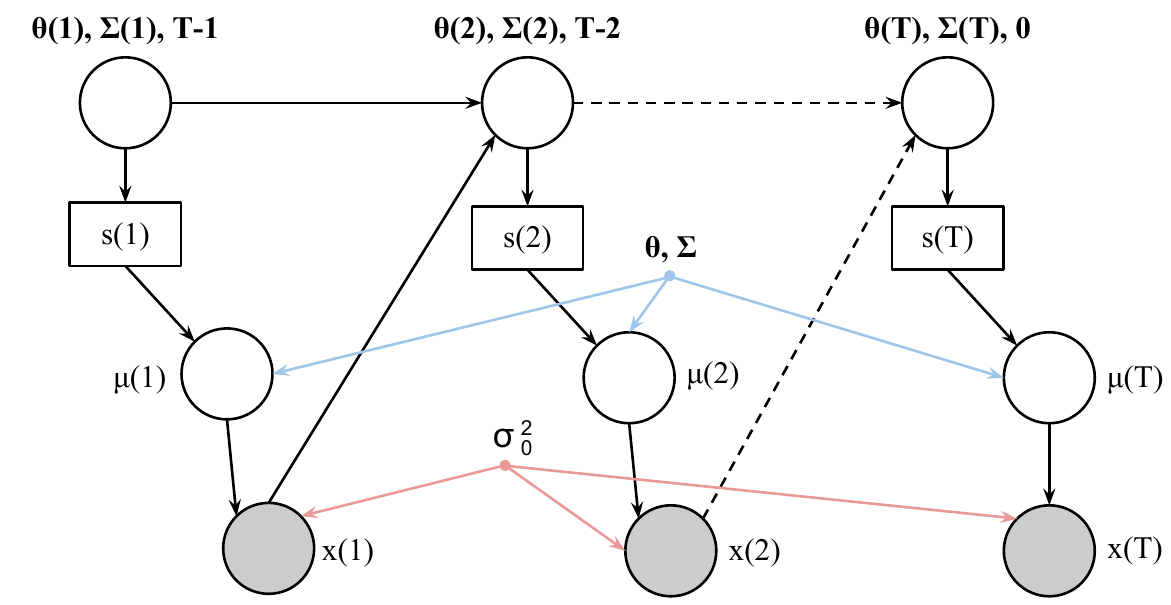}
	\caption{The payoff model illustrated by an influence diagram representation with generative processes of a finite horizon POMDP. In this diagram the non-filled circular nodes represent variables including belief states; the shaded nodes represent rewards as well as observations of the system; the black point nodes represent non-random values; and the non-filled square rectangular nodes represent actions. The dotted lines indicate indirect dependency and intermediate nodes are not drawn. Note that $s(\cdot)$ also depends on $\sigma_0^2$ but lines are not drawn for simplicity.}
	\label{fig-payoff-model-with-ad-correlation}
\end{figure}

Our goal is to choose an optimal policy to maximize $R_{\pi}(T)$. However, we do not observe any future $x_{s(t)}(t)$ (where $t\geq t'$) before making a decision about $s(t)$ (at $t=t'$).  For a given webpage, we assume two generative processes to generate the payoffs as shown in Figure \ref{fig-payoff-model-with-ad-correlation}. First, we consider the matching between ads and the webpage and denote the true but unknown payoffs of ads over a webpage as $\bs\mu$, a $N$-dimension vector. This vector is generated from a multivariate Gaussian distribution governed by mean vector  $\bs\theta$ and covariance matrix $\bs\Sigma$ as the following
\begin{equation}\label{eq-mu}
\bs{\mu} \sim \mathcal{N} ( \bs\theta, \bs\Sigma )
\end{equation}
where $\bs\theta$ and $\bs\Sigma$ are the parameters of the model and can be estimated beforehand from data. Meanwhile, considering the fact that the payoffs are affected by either the visiting users, some unexpected factors, or the uncertainty that has not been well modelled from the Gaussian, the observed payoffs are generated from the true payoffs by
\begin{equation}\label{eq-x}
\bs{X} \sim \mathcal{N} ( \bs\mu, \sigma^2_0 \cdot \bs I )
\end{equation}
where $\bs{X}$ is a $N$-dimension observed payoff vector and $\bs I$ is the identity matrix. A simple variance $\sigma^2_0$ is used to model the noise. Note that our treatment of the uncertainty from the underlying users is basic that we use a universal constant to describe the possible noise. We will show in our experiments that the noise factor, although simple, plays an important role in controlling the sensitivity of our model. For a more advanced treatment about user modelling, we refer to \cite{tang2011learning}. Given the two-stage process, the objective of the publisher is to find the optimal policy which maximizes the expectation of the overall ad income through the period, i.e.
\begin{align*}\label{eq-object-function}
 \pi^{\ast} = &\argmax_{\pi} \mathbb{E} \left[ R_\pi(T) \right] \\
=	& \argmax_{\pi} \mathbb{E} \left[ \sum_{t=1}^T X_{s(t)}(t) \right] \\
=	& \argmax_{\pi} \sum_{t=1}^T \mathbb{E} \left[ X_{s(t)}(t) \right] \\
=	&\argmax_{\pi} \sum_{t=1}^T \int_x x_{s(t)}(t) p ( x_{s(t)}(t) | \Psi(t) ) dx \\
=	& \argmax_{\pi} \sum_{t=1}^T \theta_{s(t)}(t)
	 \tageq
\end{align*}
where we drop $M$ from Equation~\ref{objectfunction} because the decision is independent of it. Note that $\theta_{s(t)}(t)$ is the shorthand for $\hat \mu_{s(t)} | \Psi(t)$, denoting the estimated expectation of $\mu_{s(t)}$ at time step $t$ giving all available information $\Psi(t)$. It presents our belief at time $t$. Our formulation is in fact a special case of continuous POMDPs \cite{cassandra1998survey, thrun2000monte}, where $\bs{\mu}$ is the hidden state and $\bs{X}$ is the observation over time and our belief about $\bs{\mu}$ would be sequentially updated using a posterior probability. Our next task is to provide the estimation $\theta_{s(t)}(t)$ at each time step $t$ after an ad has been selected and its payoff has been observed. We are particularly interested in the update for correlated ads, which will be discussed in the next section. 

\subsection{Belief Updates}

At each time step the publisher makes a decision, observes the payoff, then updates expected payoffs of all ads. In order to calculate the expected payoffs the publisher needs to calculate the value of $\bs\theta (t+1)$ and $\bs\Sigma (t+1)$ according to observation $x_{s(t)}(t)$  and previous belief. In this section we derive the update equation using Bayesian inference. Let us first look at the brief update of the same ads. Suppose the publisher has two ad candidates. The 1st ad was selected at time step $t$ and received a payoff of $x_1(t)$. With Bayes' theorem and marginalizing $\mu_1$ out, we obtain the p.d.f. of $X_1$ conditioned on the new observation $x_1(t)$ and previous available information $\Psi(t)$:
\begin{align*}\label{eq-bayes-x-1}
& p \left( x_1 | x_1(t), \Psi(t) \right) \\
	&= \int p \left( x_1 | x_1(t), \Psi(t), \mu_1 \right) p \left( \mu_1 | x_1(t), \Psi(t) \right) d\mu \tageq,
\end{align*}
where we know
\begin{align*}
&\  p \big( \mu_1 | x_1(t), \Psi(t) \big) \propto \ p( x_1(t) | \mu_1, \Psi(t) ) p( \mu_1 | \Psi(t) ) \\
\propto	&\ \exp \left\{ - \big( x_1(t) - \mu_1 \big)^2 - \big( \mu_1 - \theta_1(t) \big)^2 \right\} \tageq
\end{align*}
and by inspecting the exponent part, we can find the posterior distribution of $\mu_1$ as
\begin{align*}\label{eq-bayesian-update-i}
\mu_1 | x_1(t)
	&\sim \mathcal{N} \left( \theta_1(t+1), \sigma^2_1(t+1) \right) \tageq \\
\theta_1(t+1)
	&=  \frac{\sigma^2_1(t) x_1(t) + \sigma^2_0 \theta_1(t)}{\sigma^2_1(t) + \sigma^2_0}\\
\sigma^2_1(t+1)
	&= \frac{ \sigma^2_1(t) \sigma^2_0}{\sigma^2_1(t) + \sigma^2_0} 
\end{align*}
where similarly we write $\sigma^2_i(t)$ as the shorthand for $\sigma^2_i | \Psi(t)$.

Substituting the posterior of $\mu_1$ into Equation~\ref{eq-bayes-x-1}  gives the expected payoff of the selected ad as
\begin{equation}\label{eq-update-x-1}
X_1 | x_1(t), \Psi(t) \sim \mathcal{N} \left( 
	\theta_1(t+1),
	\sigma^2_0 + \sigma^2_1(t+1)
	\right)
\end{equation}
where recall that we assume the prior noise $\sigma^2_0$ is known.

\subsubsection{Correlated Ads}

In a real world situation the payoffs of ads are correlated. Similar products, using similar creative, or targeting similar potential customers will generate similar payoffs.  By considering the correlation of ads the publisher could find a more efficient way of identifying best candidates because not only the beliefs of the selected ads themselves can be updated using Equation~\ref{eq-update-x-1}, but so do the other correlated ads. Again with Bayes' theorem and marginalizing $\mu_1$ out, we obtain the p.d.f. of $X_2$ conditioned on the observation $x_1(t)$ and previous available information $\Psi(t)$
\begin{align*}\label{eq-bayes-x-2}
& p \left( x_2 | x_1(t), \Psi(t) \right) \\
	&= \int p \left( x_2 | \mu_2, x_1(t), \Psi(t) \right) p( \mu_2 | x_1(t), \Psi(t) ) d\mu_2 \tageq
\end{align*}
where 
\begin{align*}\label{eq-bayes-mu-2}
& p( \mu_2 | x_1(t), \Psi(t) ) \\
	&\propto p( x_1(t) | \mu_2, \Psi(t) ) p( \mu_2 | \Psi(t) ) \\
	&= p( \mu_2 | \Psi(t) ) \int p( x_1(t) | \mu_1, \Psi(t) ) p( \mu_1 | \mu_2, \Psi(t) ) d\mu_1 \tageq
\end{align*}

With the covariance known, we have the conditional distribution of $\mu_1$ on $\mu_2$ as
\begin{align*}\label{eq-conditional-distribution-j}
\mu_1 | \mu_2 &\sim \mathcal{N} (\theta_1 | \mu_2, \sigma^2_1 | \mu_2 ) \tageq \\
\theta_1 | \mu_2 &= \theta_1 + \frac{ \sigma_{1,2} }{ \sigma_2^2 } \left( \mu_2 - \theta_2 \right) \\
\sigma^2_1 | \mu_2 &= \sigma^2_1 - \frac{ \sigma^2_{1,2} }{ \sigma^2_2 } 
\end{align*}
where $\sigma_{1,2}$ is the covariance of $\{ \mu_1, \mu_2 \}$. Substituting them into Equation~\ref{eq-bayes-mu-2} gives
\begin{align*}\label{eq-bayesian-update-j}
\mu_2 | x_1(t) &\sim \mathcal{N}( \theta_2(t+1), \sigma^2_2(t+1) ) \tageq \\
\theta_2(t+1)
	&= \theta_2(t) + \sigma_{1,2} \frac{ x_1(t) - \theta_1(t) }{ \sigma^2_1(t) + \sigma^2_0 } \\
\sigma^2_2(t+1)
	&= \sigma^2_2(t) - \frac{ \sigma^2_{1,2} }{ \sigma^2_1(t) + \sigma^2_0 } 
\end{align*}

Similarly substituting the posterior of $\mu_2$ from Equation~\ref{eq-bayesian-update-j} to Equation~\ref{eq-bayes-x-2} we obtain the expected payoff of the non-selected ad as
\begin{align*}
X_2 | x_1(t), \Psi(t) \sim \mathcal{N} \left( \theta_2(t+1), \sigma^2_0 + \sigma^2_2(t+1) \right) \tageq
\end{align*}

Note the correctness of the equation can be verified by  
setting the 1st and 2nd ads equal -- if we have $\theta_1 = \theta_2$ and $\sigma^2_1 = \sigma^2_2 = \sigma_{1,2}$, then Equation~\ref{eq-bayesian-update-j} becomes exactly Equation~\ref{eq-bayesian-update-i}. 
Therefore we take Equation~\ref{eq-bayesian-update-j} as the unified equation covering both self and correlated updates. The objective function in Equation~\ref{eq-object-function} is now completed with the belief update formulas (constraints), all of which are now summarized together as the following
\begin{align*}\label{eq-object-function-final}
 \pi^{\ast} =& \argmax_{\pi} \sum_{t=1}^T \theta_{s(t)}(t) \tageq
\end{align*}
subject to
\begin{align*}
\theta_{s(t+1)}(t+1)
	&= \theta_{s(t+1)} (t)+ \sigma_{s(t), s(t+1)} 
	\frac{  x_{s(t)}(t) - \theta_{s(t)}(t) }{ \sigma^2_{s(t)}(t) + \sigma^2_0 } \tageq \label{cf} \\
\sigma^2_{s(t+1)}(t+1)
	&= \sigma^2_{s(t+1)}(t) - 
	\frac{ \sigma^2_{s(t), s(t+1)} }{ \sigma^2_{s(t)}(t) + \sigma^2_0 } \tageq \label{cf-variance}
\end{align*}

It is worth noticing that the update in Equation~\ref{cf} is closely related to the ``word of mouth'' heuristic adopted in collaborative filtering \cite{Breese:1998, Herlocker:1999}. In the collaborative filtering approaches, particularly, the user-based ones, the rating of a target user is estimated by looking at other similar users; The more similar a user is, the more contribution he or she would have to the prediction. Using the heuristic, the final rating prediction is a weighted average across all similar users \cite{Herlocker:1999} and the similarity is usually measured by cosine similarity or Pearson's correlation coefficient and user means are used to remove the bias of mean ratings among users \cite{Breese:1998}. In this paper, using a simple Gaussian model, we naturally derive a collaborative update mechanism across correlated ads. The major difference is that the update is in a sequential way. As seen in Equation~\ref{cf}, the similarity measure here is the correlation normalized by the variance, and $\theta_{s(t+1)} (t)$ and $\theta_{s(t)}(t)$ are used to remove the bias from the mean payoffs between different ads. Moreover, Equation~\ref{cf-variance} naturally provides the confidence of the predictions. 

Optimizing Equation~\ref{eq-object-function-final} leads to the exploration and exploitation dilemma. The publisher would like to earn more by selecting the best known $\theta_{s(t)} (t)$ so far. But some ads with higher variances might potentially have higher payoffs and are also required to be selected in order to gather the feedback. Not selecting the local best may result in a loss of the immediate reward, but the loss, however, could be compensated if any better alternatives can be found in later stages. Besides, we can see from the model that the changing dataset problem is dealt with naturally: the new coming ads could be simply assigned high variances to encourage the exploration on them.

\subsection{Exact Solution} 

\subsubsection{Value Iteration}

Our revenue optimization problem in Equation~\ref{eq-object-function-final} could be solved exactly following a value iteration approach using Dynamic Programming \cite{Bertsekas:1995}. Recall we use $\bs\theta$ and $\bs\Sigma$ to denote the belief of $N$ ads true payoffs before $t=1$. Let $V^{\ast}( \bs\theta, \bs\Sigma, T )$ denote the max possible revenue the publisher could gain in $T$ time steps. We have the following statement and the Bellman equation \cite{Bertsekas:1995}.
\newtheorem{lemma}{Lemma}
\begin{lemma}\label{lemma-1}
For any given priori $\bs\theta$ and $\bs\Sigma$, there exists an optimal policy $\pi^{\ast}$ to the problem in Equation~\ref{eq-object-function-final}, and $V^{\ast}( \bs\theta, \bs\Sigma, T )$ is achievable. More over $V^{\ast}( \bs\theta, \bs\Sigma, T )$ satisfy the following condition
\end{lemma}
\begin{align}\label{eq-lemma-1}
  & V^\ast \left( \bs\theta, \bs\Sigma, T \right) \notag \\
  &= \max_{s(1) \in N} \mathbb{E} \left[ X_{s(t)}(1)
	+ V^\ast \left( \bs\theta | X_{s(t)}(1), \bs\Sigma | X_{s(t)}(1),  T-1 \right) \right]
\end{align}

By solving this equation recursively we find the optimal policy. If $T=1$, we write the optimal revenue as
\begin{align*}
V^{\ast}( \bs\theta, \bs\Sigma, 1 ) &= \max_{s(1) \in N} \mathbb{E} \left[ X_{s(1)}(1) \right] \\
	&= \max_{s(1) \in N} \theta_{s(1)}(1) \tageq
\end{align*}
which indicates that the publisher should simply choose the ad with highest expected payoff. This choice is straightforward because no more time steps exist, thus no need of exploration.

For $T=2$ the optimal revenue is written as
\begin{align*}\label{eq-step-2-1}
& V^{\ast}( \bs\theta, \bs\Sigma, 2 ) \\
	&= \max_{s(1) \in N} 
		\mathbb{E} \left[ X_{s(1)}(1) + V^{\ast} \left( \bs\theta | X_{s(t)}(1), \bs\Sigma | X_{s(t)}(1), 1 \right) \right] \\
	&
\small{
	= \max_{s(1) \in N} \int p(x_{s(1)}) \left( x_{s(1)} + U^{\ast}( \bs\theta | x_{s(t)}(1), \bs\Sigma | x_{s(t)}(1), 1) \right) dx_{s(1)} 
} \\	
	&= \max_{s(1) \in N} \int p(x_{s(1)}) \left( x_{s(1)} + \max_{s(2) \in N} \left( \theta_{s(2)}(2) \right) \right) dx_{s(1)} \\
	&= \max_{s(1) \in N} \left( \theta_{s(1)}(1) + \int \max_{s(2) \in N} p(x_{s(1)}) \theta_{s(2)}(2) dx_{s(1)} \right) \tageq
\end{align*}

The difficulty lies in the last integral as it depends on the $\max$ operator. Using Chasles relation, it could be expanded to several regional integrals according to the random vector $\bs\theta(2)$. For instance, if the publisher has to choose from only two ads, and by solving $\theta_1(2) > \theta_2(2)$
the following answer is obtained
\begin{equation}
\theta_1(2) > \theta_2(2) \text{ when } x_i(1) > k
\end{equation}
which indicates the publisher should choose the 1st ad when the observation from the 1st time step is bigger than some value $k$. Then the last integral of Equation~\ref{eq-step-2-1} could be broken into two regional integrals with exact solution
\begin{align*}\label{eq-step-2-2}
V^{\ast}( \bs\theta, \bs\Sigma, 2 ) 
	&= \max_{i \in 1,2} \Big(
		\theta_1(1) + \int_{-\infty}^{k} p(x_i) \theta_2(2) dx_i \\
	&\;\;\; + \int_{k}^{+\infty} p(x_i) \theta_1(2) dx_i
	\Big) \tageq
\end{align*}

The above equation could be easily extended to $N$ ads cases provided the solution of $N$ inequalities for vector $\bs\theta$, where the only variable considered as unknown is the observation from the last time step. For simplicity we define that a region is \textit{dominated} by some ad when the ad should be selected if the observation falls in the region. This is similar with \cite{spaan2005perseus} where the value function is expressed as a linear combination of $\alpha$-functions. Formally we denote the region dominated by $i$-th ad at time step $t$ as $[m_{i,t}, n_{i,t}]$. For a general case of $N$ ads, Equation~\ref{eq-step-2-1} is written as
{
\begin{align*}\label{eq-step-2-3}
  V^{\ast}( \bs\theta, \bs\Sigma, 2 )
	&= \max_{s(1) \in N} \Big(
		\theta_{s(1)}(1) + \sum_{s(2)}^N \int_{m_{s(2),2}}^{n_{s(2),2}} p(x_i) \theta_{s(2)}(2) dx_i 
	\Big) \tageq
\end{align*}
}
where the regional integral could be solved as
{
\begin{align*}
\int_m^n x p(x) dx 
	&= - \sigma^2 \Big( p(n) - p(m) \Big) 
	+ \mu \Big( \phi(n) - \phi(m) \Big) \tageq
\end{align*}
}
where $\phi(x)$ is the c.d.f. for the Gaussian random variable $X$. Note that for some ads at some time steps, their dominant regions could be empty, simply indicating the ads would never be selected under such circumstances.

\subsubsection{A Toy Example}

We give an example to demonstrate the sequential selection mechanism with embedded correlated belief update. Assume a publisher have 2 ads to select from. The Gaussian noise from users is given by 
\begin{equation*} 
\sigma^2_0 = 0.1 
\end{equation*}

Random state $\bs\mu$  is defined by a bivariate Gaussian, i.e. 
\begin{align*}
\bs{\mu} \sim \mathcal{N} ( \bs{\theta}, \bs{\sigma} )
\end{align*}
where
\begin{align*}
\bs{\theta} = \left[
	\begin{array}{l}
	1 \\
	0.95
	\end{array}
	\right] \;
\bs\Sigma = \left[
	\begin{array}{l l}
	10 & 0.2 \\
	0.2 & 50
	\end{array}
\right]
\end{align*}

Considering only one time step gives the expected revenue
\begin{align*}
V^{\ast}(\bs\theta, \bs\Sigma, 1) &= \max_{s(1) \in N} \mathbb{E} \big[ X_{s(1)}(1) \big] \\
	&= \max (1, 0.95) \\
	&= 1 \ \	\text{with } s(1) = 1
\end{align*}

Now consider $T=2$. Suppose the 1st ad at the 1st time step is selected, which yields the following update for time step 2
\begin{align*}
\text{when }s(1) &= 1: \\
\theta_1(2) &= 1 - \frac{10 \times (1 - x_1(1) )}{0.1 + 10} \\
\theta_2(2) &= 0.95 - \frac{0.2 \times (1-x_1(1)) }{0.1 + 10}
\end{align*}
which gives
\begin{equation*}
\theta_1(2) > \theta_2(2) \text{ when } x_1(1) > 0.95
\end{equation*}

Thus, the optimal reward for $t=2$ when choosing $s(1)=1$ could be derived as
\begin{align*}
&V^{\ast}_{s(1)=1}(\bs\theta, \bs\Sigma, 2) \\
	=& 1 + \int_{0.95}^{+\infty} p(x_1) \theta_1(2)dx_1  + \int_{-\infty}^{0.95} p(x_1) \theta_2(2)dx_1 \\
 =& 3.1993
\end{align*}
A small defect is that variable $x$ stands for the payoff observed from the 1st time step and it cannot be smaller than zero.  However, due to the Gaussian assumption, when calculating $V^\ast$, we integrate $x$ over the Real space and sometime result in a negative expected reward. But such policy would vanish in later comparisons due to its low value and has little effect on our decision process.

Similarly selecting the 2nd ad at the 1st time step yields the following update for step 2
\begin{align*}
\text{when }s(1) &= 2: \\
\theta_1(2) &= 1 - \frac{0.2 \times (0.95 - x_2(1) )}{0.1 + 50} \\
\theta_2(2) &= 0.95 - \frac{50 \times (0.95 - x_2(1)) }{0.1 + 50}
\end{align*}
which gives
\begin{equation*}
\theta_1(2) > \theta_2(2) \text{ when } x_2(1) < 1.00
\end{equation*}

Thus we have
\begin{align*}
&V^{\ast}_{i=2}(\bs\theta, \bs\Sigma, 2) \\
=	&0.95 + \int_{-\infty}^{1} p(x_2) \theta_1(2)dx_2 
		+ \int_{1}^{+\infty} p(x_2) \theta_2(2)dx_2 \\
=	& 4.7291
\end{align*}

Finally the maximum expected payoff over two time steps is
\begin{align*}
V^{\ast}(\bs\theta, \bs\Sigma, 2) = \max (3.1993, 4.7291) = 4.7291
\end{align*}
and the corresponding optimal policy is
\begin{equation*}
\pi^{\ast} = \left\{ s(1)=2, s(2)=1 \text{ if } x_1(1) < 1, s(2)=2 \text{ otherwise} \right\}
\end{equation*}

From the result, we can see that the optimal policy is different from a myopic one, which tries to maximize the immediate reward only. Following the myopic policy the publisher would choose $s(1)=1$ and receive a smaller payoff. One of the reasons of selecting the 2nd is because it has a higher variance (taking into account the correlations as well).  We will show in the experiments that such high variance is ordinary in real world data.

\subsection{Approximate Solution}

In order to understand approximations we write the objection function as a combination of immediate reward estimation and exploration function, which utilizes the available information up to the decision time step, denoted by $\xi( \Psi(t), i )$ for the $i$-th candidate. The decision criteria is to maximize some the objective value function $V_{s(t)}( \Psi(t) )$, i.e. 
\begin{align}
  s(t) &= \argmax_{s(t)\in N} V_{s(t)}( \Psi(t) ) \notag \\
	&= \argmax_{i \in N} \left( \bar{x}_i + \xi( \Psi(t), i ) \right)
\end{align} 

For example, the exploration function in Equation~\ref{eq-lemma-1} is
\begin{align}
\xi = \int p(x_{s(1)}) V^\ast \left( \bs\theta | x_{s(t)}(1), \bs\Sigma | x_{s(t)}(1),  T-1 \right) dx
\end{align}
of which the computation is expensive due to recursive calling and integral. In this section we present two approximate methods.

\subsubsection{Value Iteration with Monte Carlo Sampling}

\begin{algorithm}[!t]                      
\caption{The \textsc{vi-cor} algorithm using value iteration with Monte Carlo sampling.}          
\label{algo-mc}                           
\begin{algorithmic}
\Function{ValueFunc}{$ \bs\theta, \bs\Sigma, t$}
	\State $\text{array} \; V \leftarrow 0$
	\Comment{\small Expected reward vector. \normalsize}
		
	\Loop{$\; i \leftarrow 1$ to $N$}	
		\State $V[i] \leftarrow \theta_i(t)$
		\Comment{\small Expected immediate reward. \normalsize}
		\If{$t < T$}
			\ForAll{$\; s \text{ in } \textsc{Sample}(\bs\theta, \bs\Sigma)$}
				\State $[\bs\theta', \bs\Sigma'] \leftarrow \textsc{UpdateBelief}(\bs\theta, \bs\Sigma, s, i)$
				
				\Comment{{\small New belief after selecting $i$ and observing $s$.}}
				
				\Comment{{\small Equations~\ref{cf} and \ref{cf-variance}.}}
				\small \State $V[i] \leftarrow V[i] + \frac{1}{M_0} \textsc{ValueFunction}( \bs\theta', \bs\Sigma', t+1 )$ \normalsize
			\EndFor
		\EndIf
	\EndLoop
	\State \Return $[ \textsc{Max}(V), \textsc{MaxIndex}(V) ]$
\EndFunction
\end{algorithmic}
\end{algorithm}
\begin{algorithm}[!t]                      
\caption{The \textsc{ucb1-normal-cor} algorithm using multi-armed bandit with correlated update.}          
\label{algo-ucb1-normal-cor}                           
\begin{algorithmic}
\Function{Plan}{$ \bs\theta, \bs\Sigma, \Psi(t) $}
	\State $\text{array} \; V \leftarrow 0$
	\Loop{$\; i \leftarrow 1 \text{ to } N$}
		\If{$t_i < \lceil 8 \log t \rceil$}
			\State \Return i
		\EndIf
	\EndLoop

	\State $[\bs\theta', \bs\Sigma'] \leftarrow \textsc{UpdateBelief}( \bs\theta, \bs\Sigma, \Psi(t) )$		
	
	\Comment{{\small New belief of all ads with all available information.}}
	
	\Comment{{\small Equations~\ref{cf} and \ref{cf-variance}.}}
				
	\Loop{$\; i \leftarrow 1$ to $N$}
		\State $V[i] \leftarrow \theta'_i + \sqrt{16 \cdot \frac{ q_i - t_i \theta'^2_i }{ t_i - 1} \cdot \frac{ t-1 }{ t_i }}$
		
		\Comment{\small Expected reward. \normalsize}
	\EndLoop
	\State \Return $[ \textsc{max}(V), \textsc{MaxIndex}(V) ]$
\EndFunction
\end{algorithmic}
\end{algorithm}

For $T \geq 3$ the solution for $N$ inequalities cannot be obtained easily. Instead we use Monte Carlo sampling to deal with the integral and avoid solving the inequalities. The exploration function is written as
\begin{align}\label{eq-vi-cor-exploration-func}
\xi_{\textsc{vi-cor}} 
	= \frac{1}{M_0} \sum_{x \in \mathbb{S}} p(x) 
		V^\ast \left( \bs\theta | x, \bs\Sigma | x,  T-1 \right) dx
\end{align}
where $\mathbb{S}$ is the sample set and $M_0$ is the normalizing factor from sampling. The algorithm for a general $T \geq 3$ case is represented in Algorithm~\ref{algo-mc}. We refer to this algorithm as \textsc{vi-cor} in our experiments.

\subsubsection{The {\small UCB1-Normal-COR} Algorithm}

The multi-armed bandit is a popular approach dealing with exploration-exploitation dilemma in sequential optimization problems \cite{auer2002finite}. Similar to the problem discussed above, in the multi-armed bandit scenario a player must decide which arm to play at each time step to maximize the cumulative reward over the entire planning horizon. 

Most multi-armed bandit algorithms reduce the computational cost by approximating the exploration function. In this paper we base on the deterministic policy \textsc{ucb1-normal} \cite{auer2002finite} and improve its performance by adding the correlated update. The original algorithm assumes the Gaussian distribution of the reward, independence between arms, and underlying mean and variance for reward distribution are unknown but fixed. The exploration function of \textsc{ucb1-normal} is written as
\begin{equation}\label{eq-ucb1-normal-exploration-func}
\xi_{\textsc{ucb1-normal}} 
	= \sqrt{16 \cdot \frac{q_i - t_i \theta^2_i(t)}{t_i - 1} \cdot \frac{t-1}{t_i}}
\end{equation}
where $q_i$ is the sum of squared reward obtained from $i$-th arm, and $t_i$ the times $i$-th has been played so far. We extend the algorithm for our problem by adding the correlated updates. The exploration function would remain the same form, but $\bs\theta(t)$ at each time step is updated according to Equations~\ref{cf} and \ref{cf-variance}, instead of only updating the selected candidate. The algorithm, referred to as $\textsc{ucb1-normal-cor}$ in our experiments, is represented in Algorithm~\ref{algo-ucb1-normal-cor}.

\section{Empirical Experiments}
\label{sec-experiment}

\subsection{Dataset}



We collected our test data set from Google AdWords service \cite{google-adwords-traffic-estimator}. We consider the scenario that advertisers deploy campaigns through an advertiser (demand) side platform, whereas online publishers retrieve ads and earn revenue from a related publisher side platform. Generally, online publishers share the ad revenues with their chosen ad networks or exchanges with a fixed percentage. For instance, with Google AdSense online publishers gain 68\% of advertisers' spending, where the ratio has not changed since 2003 \cite{google-adsense-share}. Thus, it is reasonable for us to consider online publishers' ad revenue to be proportional to the corresponding revenue of Google AdWords and also the corresponding cost of the advertisers.

The test data was collected from 12/2011 to 5/2012\footnote{The dataset is publicly available at \url{http://www.computational-advertising.org}.}. The Google AdWords Traffic Estimation service provides real-time data to help advertisers to adjust budgets and select appropriate keywords. When given a keyword, budget, and various targets, the service will return a list of fields on a daily basis including clicks, global and local impressions, average position, average cost-per-click, and total cost. In addition, we also separated the US and UK markets using the geographical targeting option of the service. We have collected keywords across the Google Sponsored Search and Display Networks and do not make a distinction between the types of ads and their specific payment schemes. This allows online publishers to fully concentrate upon the revenue optimisation task. During the collection period 521 different keywords from various categories were collected and 310 have non-zero mean payoffs. As shown in Table~\ref{tbl-keyword-category}, we manually categorized the keywords into 8 categories to reflect that fact that it is not necessary for online publishers to try out all the available ads; instead they should specify their target categories (based on hosting webpages content) and make optimal decisions within the targeted categories.  In our experiments, ads associated with the collected keywords were considered as candidates and decision making was on a daily basis. 



\subsection{Baselines and Experiment Setup}

\begin{table}[t]
\centering
\scriptsize
	\begin{tabular}{|l|l|l|l|}        \hline	
        \textbf{Category}   & \textbf{Count} & \textbf{Category}       & \textbf{Count} \\ \hline
        Education            & 34              & Person  \ \&  organization & 29             \\ 
        Shopping       & 86              & (Personal) finance      & 51             \\ 
        Product  \ \&  service & 20             & Information             & 52             \\ 
        Medical             & 25              & Uncategorised           & 13              \\
        \hline 
    \end{tabular}	
	\vspace{-10px}
\scriptsize
	\caption{Categorisation of the collected 310 keywords with non-zero mean payoffs.}
\label{tbl-keyword-category}
\end{table}

The baselines used in our experiment are:
\begin{compactitem}
\item \textsc{random} policy, which selects candidates randomly (uniform);

\item \textsc{myopic} policy, which selects candidates based on immediate reward estimated so far;

\item \textsc{ucb1} policy, which assumes independent between candidates and is model-free of reward distribution \cite{auer2002finite}; and

\item \textsc{ucb1-normal} policy, which assumes independent between candidates and the reward following Gaussian distribution;

\end{compactitem}

And our algorithms are:
\begin{compactitem}
\item \textsc{vi-cor} policy, which uses value iteration with Monte Carlo sampling (Algorithm~\ref{algo-mc}); and

\item \textsc{ucb1-normal-cor} policy, which consider the dependencies between candidates (Algorithm~\ref{algo-ucb1-normal-cor}).
\end{compactitem}

For each candidate we have about $150$ daily payoff data points (some candidates have less due to later start of collection). For each category (except \textit{Uncategorised}) we select candidates with close mean payoffs to form a dataset, emphasizing the challenge of planning with matching candidates. In order to test the statistical significance of algorithms performance, we divide daily payoff time series into 8 chunks for each dataset (with overlap). For each chunk, we use $20\%$ as the training set to get the prior belief of ad performances, i.e. $\bs\theta(0)$ and $\bs\Sigma(0)$. Then we go over the remaining $80\%$ with each algorithm reporting the cumulative reward at each time step. Besides, we report the averaged results with Wilcoxon signed-rank test for the significance of the best algorithm outperformed the second best within each dataset.


\subsection{Results and Discussions}

\begin{table*}[t]
\centering
    \begin{tabular}{ | l | l | l | l | l || l | l | }
        \hline
        \textbf{Datasets}                     & \textsc{myopic} & \textsc{random} & \textsc{ucb1}  & \textsc{ucb1-normal} & \textsc{vi-cor} & \textsc{ucb1-normal-cor} \\ \hline
        Education             & 21.9    & 23.0    & 30.9  & 30.9   & \textbf{41.2*}     & 27.6            \\ 
        Finance-1             & 38.5   & 27.8       & 40.9  & 26.4   & \textbf{44.5}     & 27.4            \\ 
        Finance-2             & 22.1   & 16.5      & 30.6  & 22.8   & \textbf{38.0*}     & 22.9            \\ 
        Information           & 14.1    & 12.9       & 27.8  & 15.9    & \textbf{29.4}     & 15.9             \\ 
        People  \&  Organization & 41.6   & 30.4      & 50.5  & 31.4   & \textbf{72.9*}     & 63.3            \\ 
        Shopping-1            & 17.4    & 10.6      & \textbf{42.3}  & 16.1  & 40.2      & 16.4            \\ 
        Shopping-2            & 29.9   & 14.5      & 34.3  & 75.3   & 52.9     & \textbf{79.2*}            \\ 
        Shopping-3            & 9.7    & 4.3       & 21.9  & 18.3   & \textbf{27.3}     & 19.4            \\ 
        Product  \&  Service  & 24.7   & 26.0      & 47.2  & 57.1   & \textbf{67.9*}     & 59.9            \\
        Medical				&30.5	&19.6	&52.7	&32.2	&\textbf{58.0*}	&33.5	\\
        \hline
    \end{tabular}
\caption{The overall performance comparison. The cumulative payoffs are averaged on 8 chunks then normalized w.r.t. the {\scriptsize GOLDEN} policy for a better representation. In each row the one with highest cumulative payoff is in bold and with * if the difference with the second best is significant.}
\label{tbl-results}
\vspace{10px}
\end{table*}

We compared our two algorithms with the baselines over the 10 selected datasets.  In order to compare performances across categories we normalize the cumulative revenues against the \textsc{golden} solution (always picking up the best ads) to give a better representation due to the different scales of means from different categories.  The results are summarized in Table~\ref{tbl-results} and are compared in Figure~\ref{fig-results}. We can see that, within the 10 different datasets, the proposed \textsc{vi-cor} algorithm performed the best for 8/10 with 5/8 significantly better. The \textsc{ucb1-normal-cor} algorithm performed the best for 1/10 and was significantly better in that trail. With the ``Shopping-1'' dataset, the \textsc{ucb1} algorithm performed the best, but the \textsc{vi-cor} had the comparable performance and the difference was not significant.

In Figure~\ref{fig-results} we give daily performance comparison on ``Education'' and ``People \& Organization'' datasets where algorithm were run on entire payoff data series. We cannot represent all 10 figures due to the space limit; the performances were however consistent across all datasets. We discuss our findings in the following subsections.

\subsubsection{The Importance of Exploration}

\begin{table}
{
\scriptsize
  \centering
    \begin{tabularx}{.45 \textwidth}{ |X|X|X|X|X|X| }
        \hline
        &eminem & justin bieber & michael jackson & selena gomez & wayne rooney \\ \hline
       eminem &1.00   & -0.43         & -0.58           & -0.50        & -0.73        \\ \hline
       justin bieber & ~  & 1.00          & 0.80            & 0.74         & 0.70         \\ \hline
     michael jackson  & ~  & ~          & 1.00            & 0.97         & 0.71         \\ \hline
    selena gomez   & ~  & ~          & ~            & 1.00         & 0.63         \\ \hline
   wayne rooney    & ~  & ~          & ~            & ~         & 1.00         \\
        \hline
    \end{tabularx}
\caption{The sample correlation matrix for ``People \& Organization'' category. The high correlations made the {\scriptsize UCB1} and {\scriptsize UCB1-Normal} inefficient.}
\label{tbl-cor-po}
}
\vspace{10px}
\end{table}

\begin{figure}
\centering
	\subfigure[``Education'' category, 9 candidates.]{%
	\label{fig-result-education-1}
	\includegraphics[width=.45 \textwidth, bb=0 0 7.01in 5.49in]{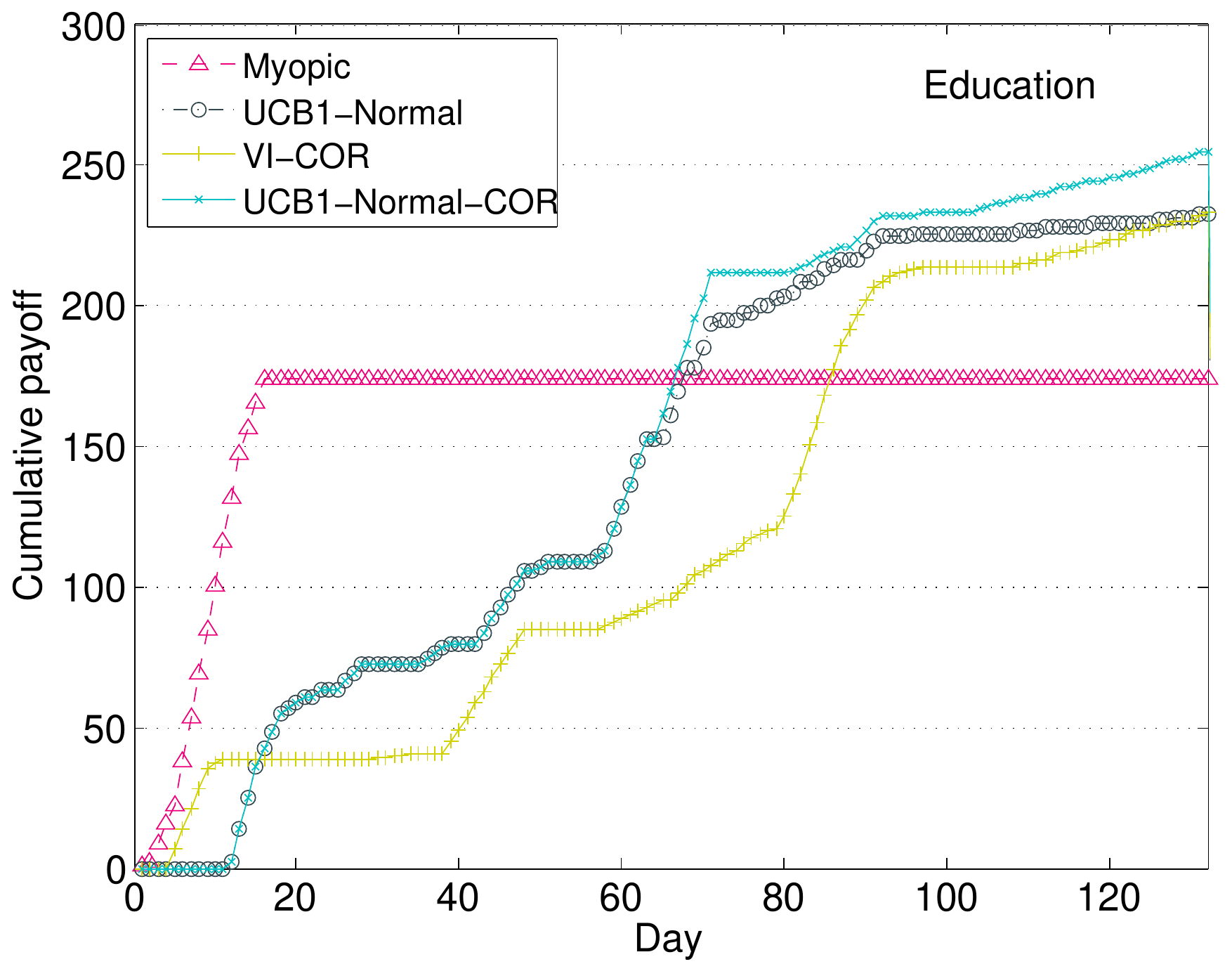}
	}%
	\vspace{-10px}
	\\
	\subfigure[``People \& organization'' category, 5 candidates.]{%
	\label{fig-result-people-organization-1}
	\includegraphics[width=.45 \textwidth, bb=0 0 7.04in 5.56in]{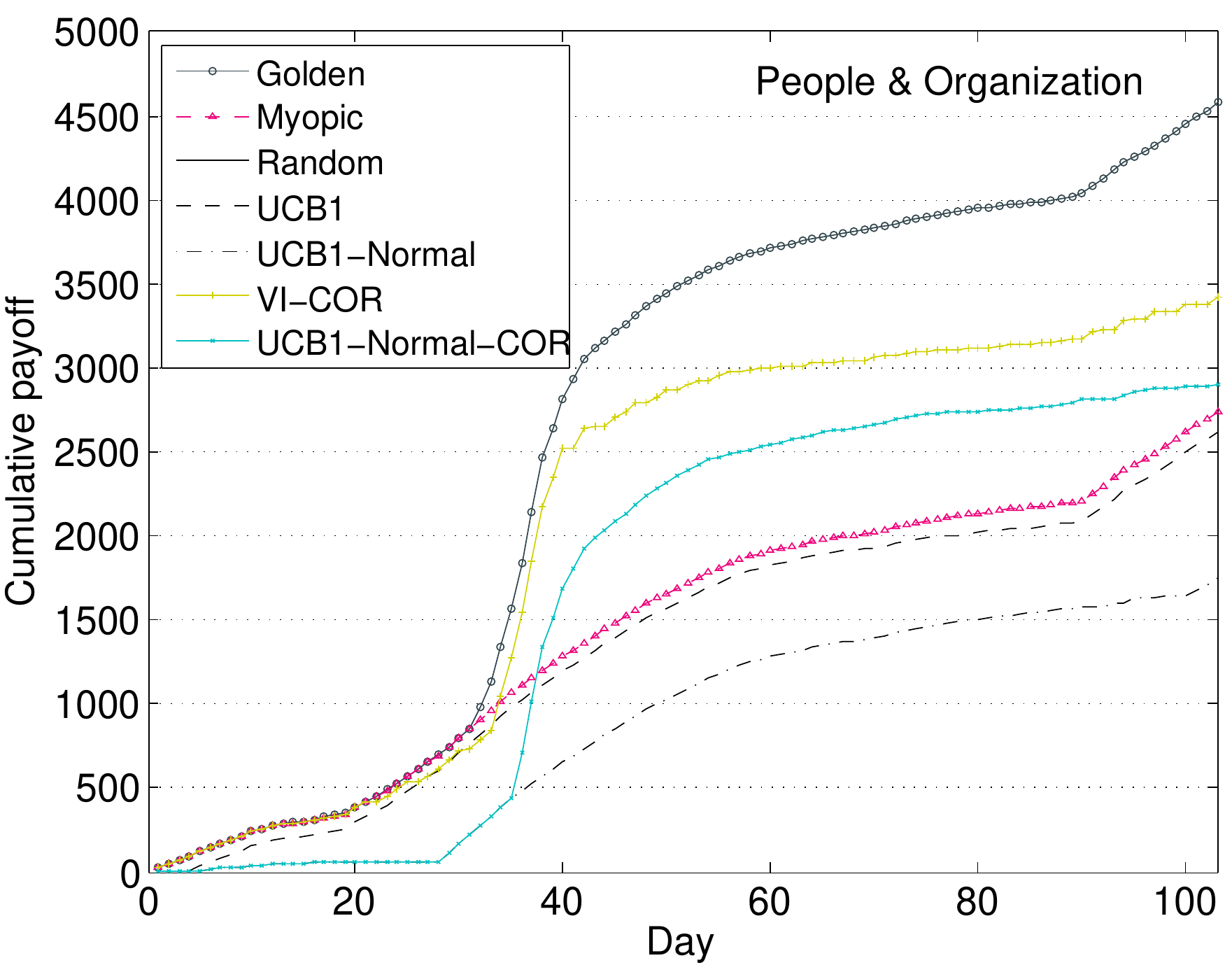}
	}%
	\vspace{-10px}
\caption{Comparison on accumulated payoffs.}
\label{fig-results}
\end{figure}

\begin{figure}[t]
	\centering
	\includegraphics[width=.45 \textwidth, bb=0 0 10.03in 6.69in]{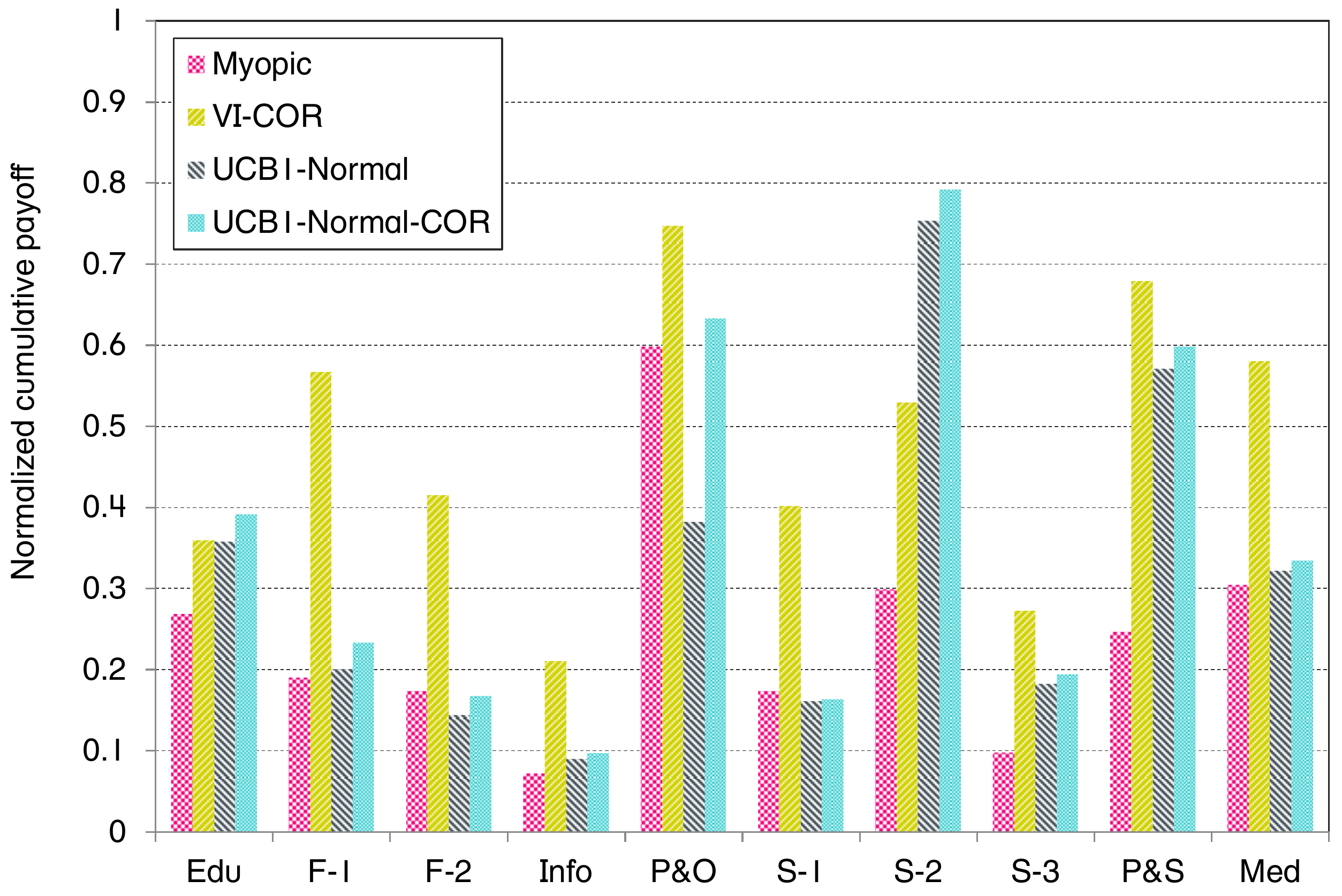}
	\vspace{-10px}	
	\caption{Comparison on accumulated payoffs on the 10 datasets. {\scriptsize VI-COR} always performed better than {\scriptsize MYOPIC} and {\scriptsize UCB1-NORMAL-COR} always performed better than {\scriptsize UCB1-NORMAL} across all datasets.}
	\label{fig-cor-comparison}
\end{figure}

Firstly we study the importance of exploration. Figure~\ref{fig-results} compares the daily accumulated revenues over time.  As illustrated in Figure~\ref{fig-result-education-1}, in the beginning (between day-0 and day-10) the \textsc{myopic} policy achieved excellent result and its cumulative payoff was the best until day-65. This is explained by the fact that there is no exploration involved in this policy and it exploits the current belief directly. However our algorithms with exploration quick outperformed it in the late stage as more profitable ads have been discovered from the exploration from the early stage. In the end, the \textsc{myopic} policy failed to win due to no exploration in the beginning, and later stuck to suboptimal ads. It is similar in Figure~\ref{fig-result-people-organization-1} where the \textsc{myopic} policy outperformed others between day-25 and day-35, but was caught up and passed very soon. These conclude that the exploration is valuable and important in the ad selection task. Note that we only show categories ``Education'' and ``People \& Organization'' only, while the algorithms behaviours were consistent across all datasets.

\subsubsection{The Effect of Correlation}


Recall we consider the Gaussian noise constant for all webpages and users. The consequence is that $X_1$ and $X_2$ are conditional independent when we know $\mu_1$ and $\mu_2$. We can further derive the relationship between covariances of $X1$ and $X_2$ and that of $\mu_1$ and $\mu_2$ as,
\begin{align*}\label{eq-covariance-derivation}
\cov[X_1, X_2]
	&= \mathbb{E} \big[ \cov[X_1, X_2 | \mu_1, \mu_2] \big] \\
	&\;\;\; + \cov \big[ \mathbb{E} [ X_1 | \mu_1, \mu_2 ], 
			\mathbb{E} [ X_2 | \mu_1, \mu_2 ] \big] \notag \\
	&= \cov[\mu_1, \mu_2] \tageq
\end{align*}
which enables publishers to use either of correlations within the model. In experiments we used $\cov[ X_1, X_2]$.


Like most multi-bandit machine approaches, the \textsc{ucb1} algorithm assumes independence between candidates. Therefore when candidates have relatively low correlations, e.g. in the ``Shopping-1'' dataset, the algorithm would perform well. This was confirmed by the observation that the \textsc{ucb1-normal-cor} algorithm reported only 0.3\% improvement over \textsc{ucb1-normal} in that dataset. But in the situation where the correlation of ads are high such as the ``People \& Organization'' category (a sample correlation matrix is shown in Table~\ref{tbl-cor-po}), our algorithms utilizing the correlation during planning showed better results. For instance the \textsc{vi-cor} algorithm showed 22.4\% improvement over the \textsc{ucb1} and the \textsc{ucb1-normal-cor} showed 31.9\% improvement over the \textsc{ucb1-normal} in the ``People \& Organization'' dataset. In Figure~\ref{fig-result-people-organization-1} it was clear that \textsc{ucb1-normal-cor} discovered the better options much quicker than the \textsc{ucb1-normal} did. The same conclusion is obtained by comparing the \textsc{vi-cor} with the \textsc{myopic} solution.  From Figure~\ref{fig-cor-comparison}, we can see a significant improvement by utilizing correlation of ads. The maximum improvement was obtained with ``Product \& Service'' (43.3\%) and on average it was 22.2\% across all experiments.

\subsubsection{The Impact of the Noise Factor $\sigma^2_0$}

\begin{figure*}[t]
\centering
	\subfigure[$\cov = 0.60$]{%
	\label{fig-cost-comparison-1}
	\includegraphics[width=.34 \textwidth, bb=0 0 6.97in 5.50in]{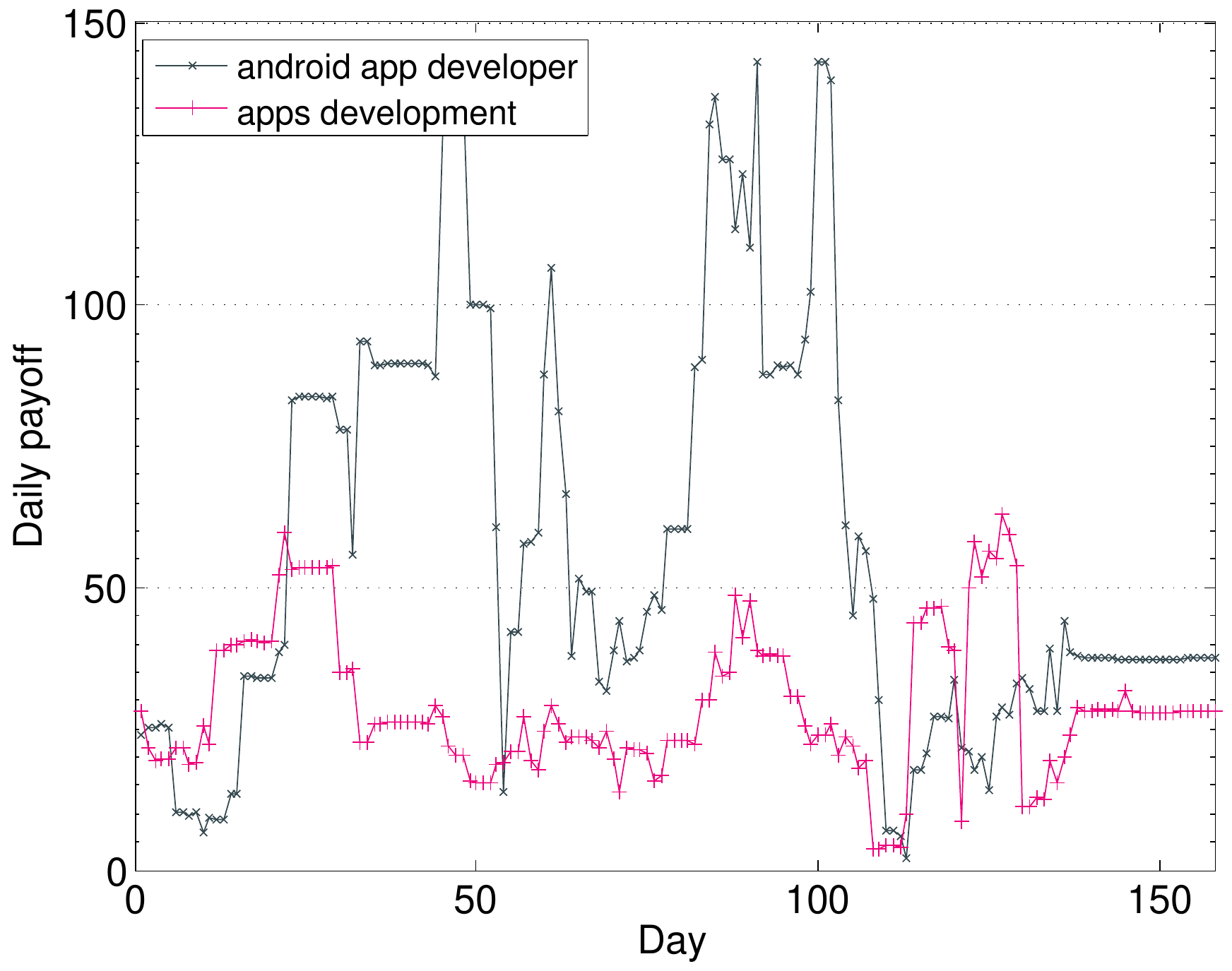}
	}%
	\subfigure[$\cov =-0.73$]{%
	\label{fig-cost-comparison-2}
	\includegraphics[width=.34 \textwidth, bb=0 0 7.08in 5.56in]{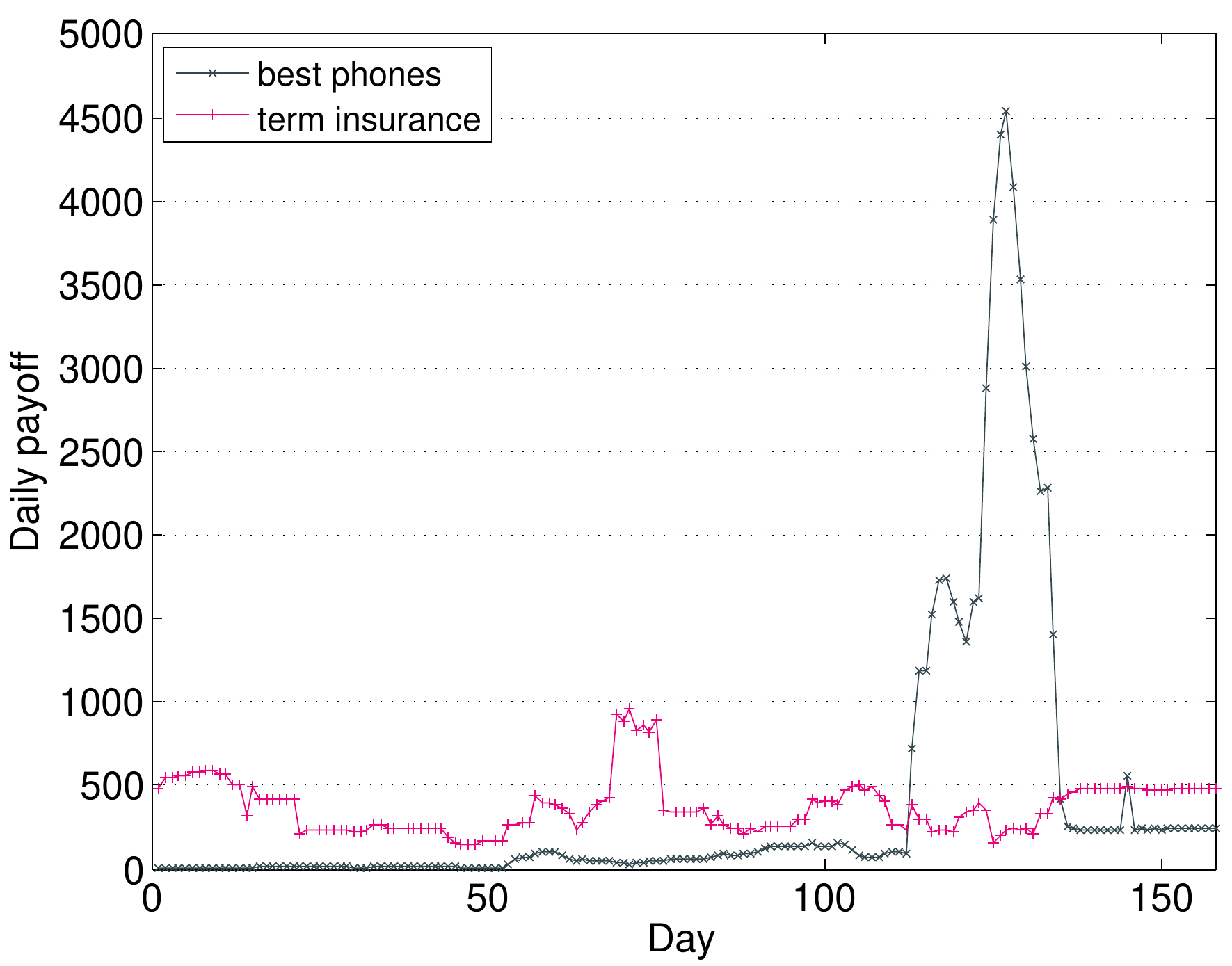}
	}%
	\subfigure[]{%
	\label{fig-noise}
	\includegraphics[width=.31 \textwidth, bb=0 0 6.68in 6.01in]{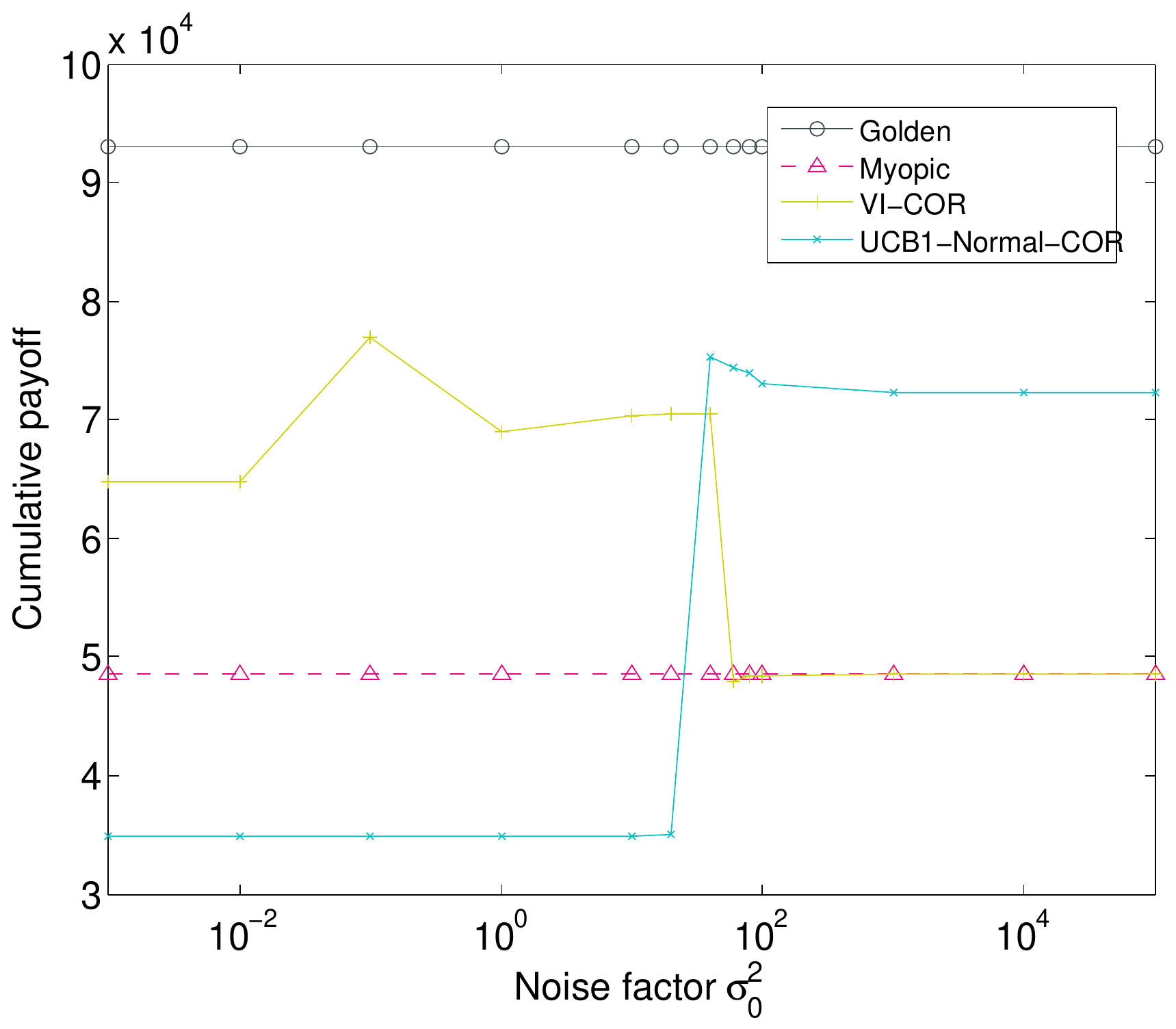}
	}%
	\vspace{-10px}
\caption{(a) The daily payoffs of two mobile development related ad candidates. (b) The daily payoff of two candidates with a sudden change. (c) The impact of the noise factor $\sigma^2_0$ for the situation in (b).}
\vspace{5px}
\end{figure*}

The introduction of the noise factor $\sigma^2_0$ is essential to the success of our model. On one hand, it helps capture the uncertainty which is unforeseeable and has not been properly modelled by the underlying $\bs\theta$ and $\bs\Sigma$ as we discussed before.  The assumption about the Gaussian distribution may not be true in practice. This can be seen from Figure~\ref{fig-cost-comparison-1} and Figure~\ref{fig-cost-comparison-2}.  The noise factor provides us with a certain flexibility of tuning our algorithms towards the specific situations. In our experiments, we obtained the noise factor by tuning using training datasets. 


On the other hand, the noise factor is also a control of the sensitivity of the algorithms towards the unexpected daily payoff fluctuation. From Equations~\ref{cf} and \ref{cf-variance} we see that the noise factor is in the denominator and a high noise factor leads to a steady policy while a low one leads to a highly responsive (sensitive) policy. A smaller value, for example $\sigma^2_0=0.01$, would make the algorithm switch probably too much, whereas a larger value would not be able to capture the fluctuation of the data. We found from the data that sometimes strong and short bursts may happen and be led by unexpected commercial activities. For instance, in the beginning of May, 2012 the Sumsang Galaxy S III and Nokia 808 PureView were presented for pre-ordering or purchasing, and both claimed to be the `best' on the market. The competition of commercial campaigns caused the daily payoff of ``best phones'' became very high between 15/04/2012 and 05/05/2012 (Figure~\ref{fig-cost-comparison-2}).  In order to response to such a short and strong abnormal activity a small noise factor is required. From Figure~\ref{fig-noise} we see that with $\sigma^2_0 \leq 40$ the \textsc{vi-cor} algorithm was able to identify and switch to ``best phones'' when the burst happened; but with $\sigma^2_0 \geq 60$ the algorithm was not able to switch, resulting in a loss of payoff.

It is worth noticing that, as illustrated in Figure~\ref{fig-noise}, the setting of the noise factor is algorithm-dependent as well. By contrast, the \textsc{ucb1-normal-cor} approach requires a large noise value to deal with the burst -- the best cumulative payoff was obtained at $\sigma^2_0=40$, and as the value of the noise factor decreased the performance dropped greatly. The different behaviour of two algorithms is due to the different structures of the exploration function. As shown in Equation~\ref{eq-ucb1-normal-exploration-func}, the exploration function of the \textsc{ucb1-normal-cor} contains the squared expectation of the payoffs in the past, indicating that the candidate with a history of low payoffs would not be favoured especially with a sudden burst. Using a high noise value would increase the chance of selecting and exploring such candidates. 

\section{Conclusions}
\label{sec-conclusion}

In this paper, we have presented a model of optimally selecting ads in an online setting. Based on POMDPs, we formulated the belief updates by taking the correlation of ads into account. We mathematically showed that the belief update across ads is similar to the ``word of month'' principle employed in collaborative filtering. Making use of the belief update, two approximate methods were proposed: one was derived from the Value Iteration and Sampling approach, whereas the other was based on the Upper Confidence Bound solution. In empirical experiments we compared our algorithms with various baselines using a collected real world dataset and showed that the Bayesian inference with correlations made the exploration more efficient and significantly improved the revenue optimization. 

In the future, we intend to extend the basic model by considering the correlations between webpages as well to further improve the exploration. A further study is also needed in order to incorporate the user click-through model and remove rank bias on the same webpage \cite{Richardson:2007}.  

{
\small
\bibliographystyle{acm}
\bibliography{bib}
}

\end{document}